\documentclass{article}
\usepackage{epsfig}

\def\be{\begin{equation}}
\def\ee{\end{equation}}
\def\bea{\begin{eqnarray}}
\def\eea{\end{eqnarray}}
\def\bc{\begin{center}}
\def\ec{\end{center}}
\def\ket#1{\hbox{$\vert #1\rangle$}}   
\def\bra#1{\hbox{$\langle #1\vert$}}   

\begin{document}

%
%

\bc
\Large{\bf NN correlations and final-state interactions in (e,e$'$NN) reactions}
\ec

\bigskip

\bc
\large Sigfrido Boffi$^{a,b}$, Carlotta Giusti$^{a,b}$, Franco Davide
Pacati$^{a,b}$, Michael Schwamb$^{a,c}$ 
\ec

\medskip

\bc
$^a$ Dipartimento di Fisica Nucleare e Teorica, Universit\`a degli Studi di
Pavia, I-27100 Pavia, Italy\\  
$^b$ Istituto Nazionale di Fisica Nucleare, Sezione di Pavia, I-27100 Pavia,
Italy\\ 
$^c$ Institut f\"ur Kernphysik, Johannes Gutenberg-Universit\"at, D-55099, Mainz,
Germany
\ec

\medskip

\noindent
After a brief overview of relevant studies on one-nucleon knockout showing the
importance of quantitatively understanding the origin of the quenched
spectroscopic factors extracted from data, attention is focussed on two-nucleon
emission as a suitable tool to investigate nucleon-nucleon correlations inside
complex nuclei. In particular, direct (e,e$'$pp) and (e,e$'$pn) reactions are
discussed, and the role of final-state interactions is studied. The influence of
the mutual interaction between the two outgoing nucleons is shown to depend on
the kinematics and on the type of the considered reaction. 

\bigskip




\section{Introduction and motivations}
Electron scattering has been used for many years as a clean tool to explore
nuclear structure. In the one-photon-exchange approximation,
where the incident electron exchanges a photon of momentum $\vec q$ and energy
$\omega$ with the target, the response of atomic nuclei as a function of
$Q^2=\vert\vec q\vert^2-\omega^2$ and $\omega$ can nicely be separated because
the electromagnetic probe and its interaction are well under control. In
addition, in direct one- and two-nucleon emission one may access the
single-particle properties of nuclei and nucleon-nucleon (NN) correlations,
respectively (see, e.g., Ref.~\cite{Oxford}).

In plane-wave impulse approximation (PWIA), i.e. neglecting final-state
interactions (FSI's) of the ejected particle, the coincidence 
(e,e$'$p) cross section in the one-photon exchange approximation is
factorized~\cite{Oxford,Report} as a product of the (off-shell) electron-nucleon
cross section $\sigma_{\rm eN}$ and the nuclear spectral density,
\be
S({\vec p}, E)=\sum_\alpha S_\alpha(E)\vert\phi_\alpha({\vec p})\vert^2.
\ee
At each removal energy $E$ the $\vec{p}$ dependence of $S({\vec p}, E)$ is given
by the momentum distribution of the quasi-hole states $\alpha$ produced in the
target nucleus at that energy and described by the (normalized) overlap
functions $\phi_\alpha$ between the target ($A$-particle) nucleus ground state and
the $(A-1)$-particle states of the residual nucleus. The spectroscopic factor
$S_\alpha$ gives the probability that such a quasi-hole state 
$\alpha$ be a pure hole state in the target nucleus. In an independent-particle
shell model (IPSM) $\phi_\alpha$ are just the single-particle states of the
model, and $S_\alpha=1$ $(0)$ for occupied (empty) states. In reality, the
strength of a quasi-hole state is fragmented over a set of single-particle
states due to correlations, and $0\le S_\alpha<1$.

Complications arise in the theoretical treatment when considering FSI's because
such a factorization in the cross section is no longer possible~\cite{Report}. 
Still, the shape of the experimental momentum distribution at each excitation
energy of the residual nucleus can be described to a high degree of accuracy 
in a wide range of kinematics in terms of quasi-hole states, and the
normalization factor needed to adjust the theoretical result to data is
interpreted as the value of the spectroscopic factor extracted from experiment. 

Two major findings came out of these studies. First, the valence quasi-hole
states $\phi_\alpha$ almost overlap the IPSM functions with only a
slight ($\sim10\%$) enlargement of their rms radius. Second, a systematic
suppression of the single-particle strength of valence states  as compared to
IPSM has been observed all over the periodic table. A quenching
of spectroscopic factors is naturally conceived in nuclear many-body theory in
terms of nucleon-nucleon correlations. However, model
calculations produce spectroscopic factors $S_\alpha$ much larger than those
extracted in low-energy (e,e$'$p) data. As an example, for
the p-shell holes in $^{16}$O a Green-function approach to the
spectral density~\cite{Polls} gives  $S_{p_{1/2}} = 0.890$ and $S_{p_{3/2}} =
0.914$, while from experiment one has $S_{p_{1/2}} = 0.644$ and
$S_{p_{3/2}} = 0.537$. In contrast, at higher energy and momentum transfer
much larger spectroscopic factors are extracted, i.e. $S_{p_{1/2}} = 0.73$ and
$S_{p_{3/2}} = 0.71$ in Ref.~\cite{Udias99}, and $S_{p_{1/2}} = 0.72$ and
$S_{p_{3/2}} = 0.67$ in Ref.~\cite{Kelly99}. This was confirmed in the
reanalysis of the $^{12}$C(e,e$'$p) data~\cite{Lapikas} where at $Q^2\le 0.3$
GeV$^2$ the s- and p-shell strength has been found quite substantially reduced
by the factor $0.57\pm 0.02$. In contrast, for $Q^2\ge 1$ GeV$^2$ the same
analysis gives a strength approximating the IPSM value. A possible
$Q^2$ dependence of spectroscopic factors, jumping from values around 0.6--0.7
at low $Q^2$ to unity at $Q^2\ge 1$ GeV$^2$, simply means that
something is not under control in either experiment or theory or both. 

In fact, the most general form of the coincidence cross section in the
one-photon-exchange approximation is the contraction of the lepton tensor
$L_{\mu\nu}$ with the hadron tensor $W^{\mu\nu}$. The latter is a bilinear form
of the hadron current $J^\mu$, i.e.
\be
J^\mu = \int d\vec{r}\ \bra{\Psi_{\rm f}}\, j^\mu(\vec{r})\,\ket{\Psi_{\rm i}}
e^{i\vec{q}\cdot\vec{r}},
\label{eq:hadron}
\ee
where the charge-current operator $j^\mu(\vec{r})$ is responsible for the
transition from an initial state $\ket{\Psi_{\rm i}}$ (describing the motion
of the ejected nucleon in its initial bound state) to a final state
$\ket{\Psi_{\rm f}}$ with the ejectile undergoing FSI's with the residual nucleus.
In the nonrelativistic PWIA approach, the representation of $\ket{\Psi_{\rm i}}$
is identified with $[S_\alpha]^{1/2}\phi_\alpha({\vec r})$ and $\ket{\Psi_{\rm
f}}$ becomes a plane wave. 

In order to get reliable information in a comparison between theory and data all
sources of theoretical uncertainties must be under control and treated
consistently. Under quasi-free kinematics $j^\mu(\vec{r})$ is reliably
approximated by a one-body operator. Ambiguities arising from its off-shell
behaviour~\cite{deForest} have been studied and shown to give a small
effect~\cite{Kelly,Udias,meucci02}. The relevance of genuine relativistic effects has
recently been investigated~\cite{Meucci} in a consistent comparison between
nonrelativistic and relativistic calculations within the distorted-wave impulse
approximation (DWIA). Significant relativistic effects,
especially in the transverse responses, are found already for a proton kinetic
energy as low as 100 MeV. As a consequence, a satisfactory description of
$^{16}$O data at low and high $Q^2$ is obtained with (extracted) spectroscopic
factors of about 0.7.

Most important is the treatment of FSI's as much of the quenching of the extracted
spectroscopic factor depends on the loss of flux introduced by FSI's in the
observed channel. A systematic analysis of the effects of FSI's is highly
desirable also in view of the debated problem of hadron propagation in the
nuclear medium and nuclear transparency. In fact, the role of genuine
attenuation of FSI's with increasing energy must be understood before studying
other mechanisms, such as e.g. colour transparency.  

Here, great help comes from the measurement of the recoil proton polarization
$P^{\rm N}$ normal to the scattering plane of the polarized incident electrons.
Without FSI's, $P^{\rm N}=0$. Therefore $P^{\rm N}$ is a good candidate to look at
when studying nuclear transparency, as its $Q^2$ dependence reflects the energy
dependence of FSI's. Relativistic DWIA results are indeed sensitive to the model
used to simulate FSI's~\cite{Meucci}.

In principle, the absorption of the ejectile is due to the same
(energy-dependent) mean field producing the quasi-hole state. The relevant
quantity is the self-energy which is obtained from a self-consistent
calculation of the nucleon spectral function. In this way it is then possible to
analyse data at different values of $Q^2$ with the same quasi-hole wave
functions and a correspondingly consistent treatment of
FSI's~\cite{rad-roth,rad-dick}. It is remarkable that with this approach the same
spectroscopic factors used to describe the data at low $Q^2$ also
describe data at high $Q^2$. Thus the puzzle on the $Q^2$ dependence seems to be
solved.

The problem remains to understand quantitatively the quenching of spectroscopic
factors with respect to IPSM and to have a handle to discriminate between
different contributions to NN correlations, ie. long/short range,
central/tensor correlations, etc. (see, e.g., Refs.~\cite{gurts,Barbieri}).
There is accumulating evidence for enhanced (e,e$'$p) transverse strength of
non-single particle origin at high missing energies~\cite{Ulmer,Dutta,Liyanage}.
However, one-nucleon emission is only an indirect tool for such a purpose. It is
by now well established that better information on NN correlations can be
obtained with exclusive two-nucleon emission. In this paper, a review of the
present status of such a type of reactions on complex nuclei is presented with
particular attention to recent work improving the treatment of
FSI's~\cite{schwamb1,schwamb2}.


\section{Two-nucleon emission}

Exclusive two-nucleon emission by an electromagnetic probe has been proposed
long time ago~\cite{Gottfried} to study NN correlations. Data with real photons
are available~\cite{McG95,Lam96,McG98,Wat00} confirming the validity of the
direct mechanism for low values of the excitation energy of the residual
nucleus. Due to the difficulty of measuring exceedingly small cross sections in
triple coincidence, only with the advent of high-duty-cycle electron beams  a
systematic investigation of (e,e$'$NN) reactions has become possible. At
present, only a few pioneering measurements have been carried
out~\cite{Onderwater97,Onderwater98,Starink,Rosner}, but the prospects
are very encouraging. 

The general theoretical framework involves the two-hole spectral
density~\cite{Boffi,GP,Oxford}, whose strength gives the probability of removing
two nucleons from the target, leaving the residual nucleus at some excitation
energy. Integrating the two-hole spectral density over the energy of the
residual nucleus one obtains the two-body density matrix incorporating
NN correlations. The triple coincidence cross section is again a
contraction between a lepton and a hadron tensor. It contains the
two-hole spectral density through bilinear products of hadron
currents $J^\mu$ of the type (\ref{eq:hadron}) suitably adapted to this type of
reaction, i.e.
\be
J^\mu = \int d\vec{r}\ \bra{\psi_{\rm f}}\, j^\mu(\vec{r})\,\ket{\psi_{\rm i}}
e^{i\vec{q}\cdot\vec{r}},
\label{eq:hadroncurr}
\ee
where $\ket{\psi_{\rm i}}$ is the two-nucleon overlap function between the
ground state of the target and the final state of the residual ($A-2$) nucleus,
and $\ket{\psi_{\rm f}}$ is the scattering wave function of the two ejected
nucleons. The nuclear current operator $j^\mu(\vec{r})$ is the sum of a one- and
a two-body part. The one-body part consists of the usual charge operator and
the convection and spin currents. The two-body part consists of the 
nonrelativistic meson exchange currents (pionic seagull and pion-in-flight
contributions) and intermediate isobar contributions such as the
$\Delta$-isobar. 

A consistent treatment of FSI would require a genuine three-body approach for
the interaction of the two emitted nucleons and the residual nucleus, which
represents a challenging task never addressed up to now in complex nuclei. A 
crucial assumption adopted in the past was the complete neglect of the mutual
interaction between the two outgoing nucleons. Thus FSI is simply described by
an attenuated flux of each ejectile due to an optical model potential. This
apparently reasonable assumption has to be checked, however. In the next
subsections first steps towards a complete description of FSI will be presented
and discussed.

Even without FSI the two-hole spectral density is not factorized in the
triple coincidence cross section. This makes a difficult task to extract
information on correlations from data, and models are required to investigate
suitable kinematic conditions where the cross section is particularly sensitive
to correlations. A priori one may envisage that two-nucleon knockout is due
to one- and two-body currents. Of course, one-body currents are only effective  
if correlations are present so that the nucleon interacting with the incident
electron can be knocked out together with another (correlated) nucleon. In
contrast, two-body currents, typically due to meson exchanges and isobar
configurations, lead naturally to two-nucleon emission even in an
independent-particle shell model. 

Two-body currents are mainly transverse and preferentially involve a
proton-neutron pair. Thus reactions like ($\gamma$,pn) and (e,e$'$pn) are
particularly sensitive to their effects. In this respect, (e,e$'$pp) reactions,
where two-body currents play a minor role, are better suited to look for
correlations. Resolution of discrete final states has been shown to provide
an interesting tool to discriminate between contributions of different
mechanisms responsible for two-nucleon emission~\cite{Giusti98}.


\subsection{NN correlations} 

The shape of the angular distribution of the two emitted nucleons mainly
reflects the momentum distribution of their c.m. total angular momentum $L$
inside the target nucleus~\cite{Boffi}. When removing, e.g., two protons from
the $^{16}$O ground 
state, the relative $^1S_0$ wave of the two protons is combined with $L=0$ or 2
to give $0^+$ or $2^+$ states of the residual $^{14}$C nucleus, respectively,
while the relative $^3P$ waves always occur combined with a $L=1$ wave function
giving rise to $0^+, 1^+, 2^+$ states. Combining the reaction description of
Ref.~\cite{GP} with the many-body calculation of the two-particle spectral
function in $^{16}$O of Ref.~\cite{Geurts}, in Ref.~\cite{Giusti98} the cross
section for the $0^+$ ground state, and to a lesser extent also for the first
$2^+$ state of $^{14}$C, was shown to receive a major contribution from the
$^1S_0$ knockout. Such transitions are therefore most sensitive to short-range
correlations. This is indeed the case, as seen in two exploratory studies
performed at NIKHEF~\cite{Onderwater97,Onderwater98}, and confirmed in
Ref.~\cite{Starink}. As the calculations are sensitive to the treatment of
correlations, precise data could give important
constraints when modelling the off-shell behaviour of the NN potential.

Superparallel kinematics has been preferred at Mainz~\cite{Rosner}, with one
proton ejected along the virtual photon direction and the other in the opposite
direction. In this kinematics only the pure longitudinal (L) and pure
transverse (T) structure functions occur in the cross section, and a Rosenbluth
L/T separation becomes possible in principle. The effect of two-body currents is
further suppressed by looking at 
the longitudinal structure function that is most sensitive to short-range
correlations. The data are still preliminary and require further analysis before
a fully reliable comparison with calculations can be done. Nevertheless they
show distinctive features predicted by calculations~\cite{Giusti98,Ryckebusch}.

Tensor correlations are expected to play a major role in (e,e$'$pn) reactions
where, however, the proton-neutron pair is ejected by a much more
complicated mechanism involving two-body currents. In the superparallel
kinematics of the proposed Mainz experiment~\cite{A1-5-98} with an incident
electron energy of 855 MeV, $\omega=215$ MeV and $q=316$ MeV/$c$ the predicted
cross sections for (e,e$'$pn) are about one order of magnitude larger than the
corresponding cross sections for (e,e$'$pp) reactions~\cite{GP99}. This
enhancement is partly due to meson-exchange 
currents and partly to tensor correlations. Quite different results are
predicted depending on these correlations being included or not. An accurate
determination of the two-hole spectral density is thus most desirable in order
to disentangle the effects of two-body currents from those of nuclear
correlations. 

Experimentally, additional and precise information will come from measurements
of the recoil polarization of the ejected proton in either (e,e$'$pp) or
(e,e$'$pn). Resolving different final states is a precise filter to disentangle
and separately investigate the different processes due to correlations and/or
two-body currents. The general formalism is available~\cite{GP00} and has been
extended to study polarization observables also in the case of two nucleons
emitted by a real photon~\cite{GP01}.


\begin{figure}[t]
\centerline{\psfig{figure=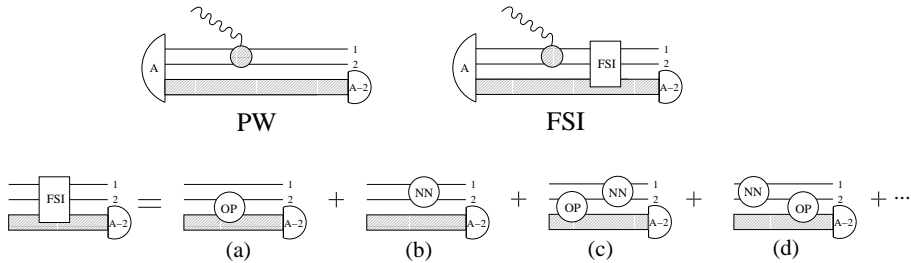,width=12cm,angle=0}}
\vspace{0.5cm}
\caption{The relevant diagrams for electromagnetic two-nucleon knockout on a
complex nucleus. The two diagrams on top depict the plane-wave approximation
(PW) and the distortion of the two outgoing proton wave functions by the final
state interaction (FSI). Below, the relevant mechanisms of FSI are depicted in
detail, where the open circle denotes either a nucleon-nucleus interaction given
by a phenomenological optical potential (OP) or the mutual interaction between
the two outgoing nucleons (NN). Diagrams which are given by an interchange of
nucleon 1 and 2 are not depicted.}
\label{fig:fig1}
\end{figure}

\subsection{Final-state interactions}

The relevant diagrams for electromagnetic two-nucleon knockout on a complex
nucleus are depicted in Fig.~\ref{fig:fig1}. In the simplest approach any
interaction between the two nucleons and the residual nucleus is neglected and a
plane-wave (PW) approximation is assumed for the two outgoing nucleons. In the
more sophisticated approach of Ref.~\cite{Giusti98}, the interaction between
each of the outgoing nucleons and the residual nucleus is considered in the
distorted-wave (DW) approximation by using a
complex phenomenological optical potential $V^{\rm OP}$ for nucleon-nucleus
scattering which contains a central, a Coulomb and a spin-orbit
term~\cite{Nad81} (see diagram (a) in Fig.~\ref{fig:fig1}). Only very recently
the mutual NN interaction $V^{\rm NN}$ between the two outgoing nucleons
(NN-FSI) has been taken into account~\cite{schwamb1,schwamb2} (diagram (b) in
Fig.~\ref{fig:fig1}). Multiscattering processes like those described by diagrams
(c) and (d) of Fig.~\ref{fig:fig1} are still neglected and left for future work.
The present treatment of incorporating NN-FSI is denoted as DW-NN. We denote as
PW-NN the treatment where only $V^{\rm NN}$ is considered and $V^{\rm OP}$ is
switched off. 

Denoting by $\ket{\vec{q}_i}$ a plane-wave state of the ejectile $i$
with momentum $\vec{q}_i$ and by $\ket{\phi^{\rm OP}(\vec{q}_i)}$ its state
distorted by the optical potential,  
the corresponding final states in these different approximations are given by
\bea
\label{fsi-pw}
\ket{\psi_{\rm f}}^{\rm PW} 
&=&
\ket{\vec{q}_1}\,\ket{\vec{q}_2}, \\
\ket{\psi_{\rm f}}^{\rm DW} 
&=&
\ket{\phi^{\rm OP}(\vec{q}_1)}\,\ket{\phi^{\rm OP}(\vec{q}_2)}, \\
\ket{\psi_{\rm f}}^{\rm PW-NN} 
&=& 
\ket{\vec{q}_1}\,\ket{\vec{q}_2}
+  G_0(z) T^{\rm NN}(z) \ket{\vec{q}_1}\,\ket{\vec{q}_2},
\label{fsi-pw-nn}
\eea
where the NN-scattering amplitude $T^{\rm NN}$ is given by 
\be
\label{tnn}
T^{\rm NN}(z) = V^{\rm NN} + V^{\rm NN} G_0(z) T^{\rm NN}(z), 
\ee
with
\be
\label{g0}
G_0(z) = \frac{1}{z-H_0(1)-H_0(2)},
\end{equation} 
$H_0(i)$ denoting the kinetic energy operator for particle $i$, and $z=
\vec{q}_1^{\,2}/(2m) + \vec{q}_2^{\,2}/(2m) + i \epsilon$. The full approach
including diagrams (a) and (b) in
Fig.~\ref{fig:fig1} gives
\be
\label{fsi-state}
 \ket{\psi_{\rm f}}^{\rm DW-NN} = \ket{\phi^{\rm OP}(\vec{q}_1)}\,
\ket{\phi^{\rm OP}(\vec{q}_2)} + G_0(z) T^{\rm NN}(z) 
\ket{\vec{q}_1}\,\ket{\vec{q}_2}.
\ee


\begin{figure}[ht]
\centerline{\psfig{figure=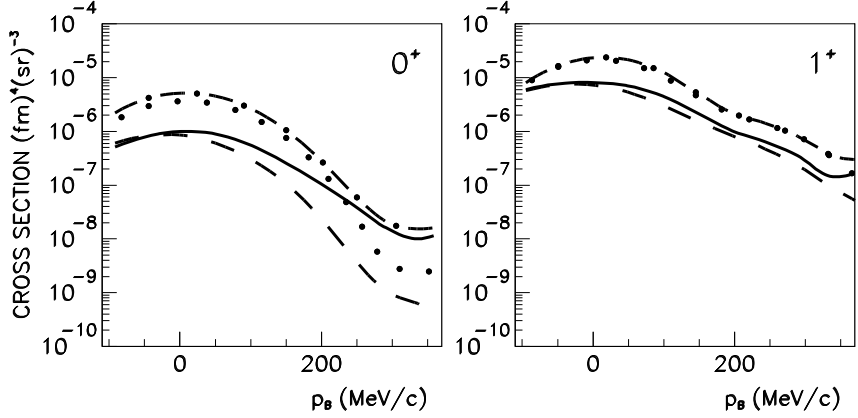,width=13cm,angle=0}}
\caption{The differential cross section of the $^{16}$O(e,e$'$pp) reaction to 
the $0^+$ ground state of $^{14}$C (left panel) and of the $^{16}$O(e,e$'$pn) 
reaction to the $1^+$ ground state of $^{14}$N (right panel) in a  
superparallel kinematics with an incident electron energy $E_0= 855$ MeV,  
an electron scattering angle $\theta_e = 18^{\circ}$, energy transfer 
$\omega=215$ MeV and  $q=316$ MeV/$c$.
In $^{16}$O(e,e$'$pn) the proton is ejected parallel and the neutron 
antiparallel to $\vec{q}$.  
Different values of $p_{\rm B}$ are obtained changing the kinetic energies of
the outgoing nucleons. Positive (negative) values of $p_{\rm B}$ refer to
situations where  
${\vec p}_{\rm B}$ is parallel (anti-parallel) to ${\vec q}$.
 Line convention:
 PW (dotted), PW-NN (dash-dotted), DW (dashed), DW-NN (solid).
}
\label{result1}
\end{figure}

\subsection{Results}

Results are presented in this section for the specific case of two-nucleon
knockout by electron scattering off $^{16}$O. Calculations have been done in the
same kinematic conditions as in previous experiments performed at
Mainz~\cite{Rosner} and NIKHEF~\cite{Onderwater97,Onderwater98}, but here
only the superparallel kinematic conditions adopted at Mainz~\cite{Rosner} will
be discussed. The differential cross sections of the $^{16}$O(e,e$'$pp)
reaction to  
the  $0^+$ ground state of $^{14}$C and of the $^{16}$O(e,e$'$pn) reaction to 
the $1^+$ ground state of $^{14}$N, calculated with the different 
approximations~(\ref{fsi-pw})-(\ref{fsi-pw-nn}) and ~(\ref{fsi-state}), are
displayed in the left and right panels of Fig.~\ref{result1}, respectively.
 
The inclusion of the optical potential leads, in both reactions, to an overall
and substantial reduction of the 
calculated cross sections (see the difference between the PW and DW results).
This effect is well known and it is mainly due to the imaginary part of 
the optical potential, that accounts for the flux lost to inelastic channels in
the nucleon-residual nucleus elastic scattering. The optical potential gives the
dominant contribution of FSI's for recoil-momentum values up to  
$p_{\rm B} \simeq 150$ MeV/$c$. At larger values NN-FSI gives an enhancement of 
the cross section, that increases with $p_{\rm B}$. In (e,e$'$pp) this 
enhancement 
goes beyond the PW result and amounts to roughly an order of magnitude for 
$p_{\rm B} \simeq 300$ MeV/$c$. In (e,e$'$pn) this effect is still sizeable
 but  much weaker.  
We note that in both cases the contribution of NN-FSI is larger in the DW-NN 
than in the PW-NN approximation.       
 
\begin{figure}[ht]
\centerline{\psfig{figure=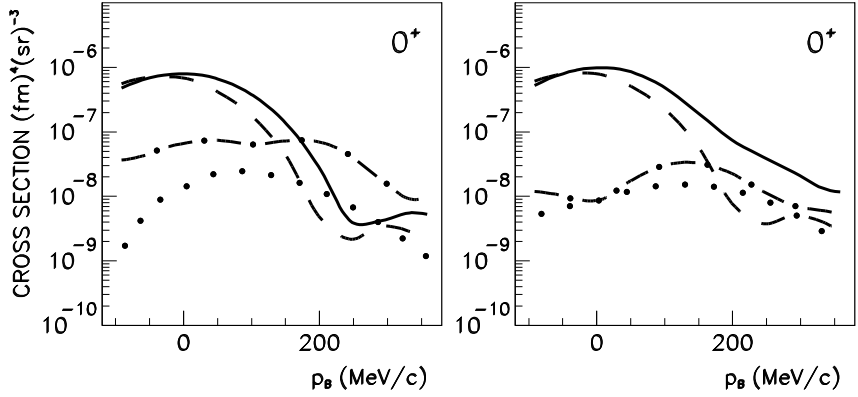,width=13cm,angle=0}}
\caption{The differential cross section of the $^{16}$O(e,e$'$pp) reaction to 
the $0^+$ ground state of $^{14}$C in the same superparallel kinematics as in
Fig.~\protect{\ref{result1}}. Line convention for the left panel:
 DW with the $\Delta$-current (dotted), 
 DW-NN with the $\Delta$-current (dash-dotted),
 DW with the one-body current (dashed),
 DW-NN with the one-body current (solid). In the right panel the dashed (solid)
curve shows the separate contribution  of the   
$^1$S$_0$ relative partial wave  in a DW (DW-NN) calculation. The dotted 
(dash-dotted) curve shows the separate contribution of the  $^3$P$_1$ relative 
partial wave in a DW (DW-NN) calculation.
}
\label{result2}
\end{figure}

In (e,e$'$pp)  NN-FSI produces a strong enhancement of the $\Delta$-current 
contribution for all the values of $p_{\rm B}$ (left panel of
Fig.~\ref{result2}). Up to about 100-150 MeV/$c$,  
however, this effect is completely overwhelmed by the dominant contribution 
of the one-body current, while for larger values of $p_{\rm B}$, where the
one-body current is less important in the cross section, the increase of the 
$\Delta$-current is responsible for the substantial enhancement in the final 
result. The effect of NN-FSI on the one-body current is
much weaker but anyhow sizeable, and it is responsible for the NN-FSI effect at
lower values of $p_{\rm B}$.
 
The combined role of FSI and the different partial waves in the initial relative
state of the two emitted protons in the $^{16}$O(e,e$'$pp) reaction is shown in
the right panel of Fig.~\ref{result2}. The effect of NN-FSI is more important on
the $^1$S$_0$ 
initial state and gives in practice almost the full contribution of NN-FSI.
The role of NN-FSI on the $^3$P initial relative states is of special
relevance for the transition to the $1^+$ excited state of $^{14}$C,  
where only $^3$P components are present and the $^1$S$_0$ relative partial 
wave cannot contribute~\cite{schwamb1}.
 
In the NIKHEF kinematics~\cite{Onderwater97,Onderwater98}, the effect of NN-FSI
is also sizeable, although not as strong as in the superparallel
kinematics~\cite{schwamb1}.
Moreover, whereas in the superparallel kinematics the relative effect of NN-FSI
increases for decreasing cross section, in the NIKHEF kinematics NN-FSI is
maximal when also the cross section is maximal, i.e. when $\vec{p}_B \approx 0$
MeV/$c$. This result clearly shows that the role of NN-FSI is strongly dependent
on the kinematics and no general statement can be drawn with respect to its 
relevance. 

\begin{figure}[ht]
\centerline{\psfig{figure= 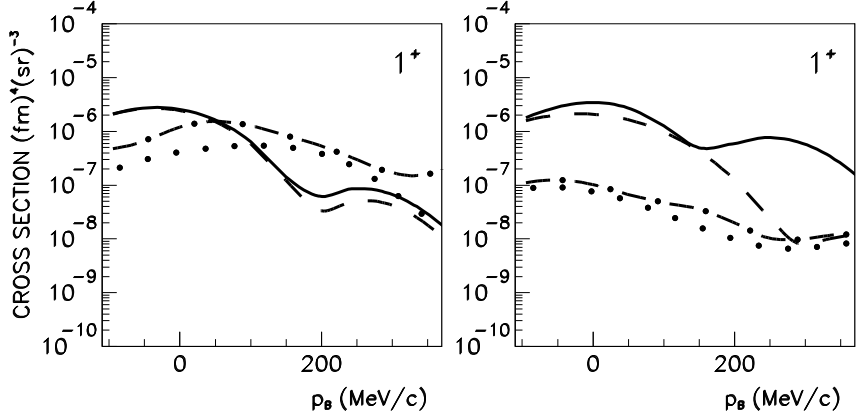,width=13cm,angle=0}}
\caption{The differential cross section of the $^{16}$O(e,e$'$pn) reaction to 
the $1^+$ ground state of $^{14}$N in the same superparallel kinematics as in
Fig.~\protect{\ref{result1}}. 
Line convention in the left panel:
 DW with the $\Delta$-current (dotted), 
  DW-NN with  the $\Delta$-current (dash-dotted),
 DW  with the one-body-part (dashed),
 DW-NN  with the one-body-part (solid).
 Line convention in the right panel:
 DW  with the pion-in-flight-current (dotted), 
 DW-NN with the pion-in-flight--current (dash-dotted),
 DW with the seagull-current (dashed),
 DW-NN with  the seagull-current (solid).
}
\label{result4}
\end{figure}

Different effects of NN-FSI on the various components of the current are
shown for the (e,e$'$pn) reaction in Fig.~\ref{result4}. Also in this
case, NN-FSI affects more the two-body than the one-body current. A sizeable
enhancement is produced on the $\Delta$-current, at all the values of 
$p_{\rm B}$, 
and a huge enhancement on the seagull current at large momenta. In contrast, 
the one-body current is practically unaffected by NN-FSI up to about 150 
MeV/$c$. A not very large but visible enhancement is produced at larger 
momenta, where, however, the one-body current gives only a negligible
contribution to the final cross section. The role of the pion-in-flight 
term, in both DW and DW-NN approaches, is practically negligible in the cross
section. Thus, a large effect is given by NN-FSI on the seagull and the 
$\Delta$-current. The sum of the 
two terms, however, produces a destructive interference that leads to a partial 
cancellation in the final cross section. The net effect of NN-FSI in 
Fig.~\ref{result1} is not large but anyhow non negligible. Moreover, the
results for the partial contributions in Fig.~\ref{result4} indicate that
in pn-knockout NN-FSI can be large in particular situations and therefore 
should in general be included in a careful evaluation. 


\section{Conclusions}

The advent of high-energy continuous electron beams coupled to high-resolution 
spectrometers has opened a new era in the study of basic nuclear properties such
as single-particle behaviour and NN correlations by means of one-
and two-nucleon emission. In parallel new theoretical approaches have been
developed. For one-nucleon knockout relativistic effects have been shown to be
most important and to affect the interpretation of data 
even at moderate energies of the emitted particles. In addition, a consistent
treatment of the initial and final states in terms of the same
(energy-dependent) Hamiltonian seems to avoid the striking feature coming
out of previous analyses of (e,e$'$p) world data with an apparent $Q^2$
dependence of the extracted spectroscopic factors. The problem remains, however,
concerning the discrepancy between calculated and observed spectroscopic
factors. This is clearly tight to NN correlations, their theoretical
treatment and the possibility of finding observables sensitive to them.

Exclusive experiments with direct two-nucleon emission by an electromagnetic
probe have been suggested long time ago as good candidates to study
correlations. In electron scattering they require triple coincidences with three
spectrometers. This is now possible and the first experiments have been
performed. By an appropriate selection of the kinematic conditions and specific
nuclear transitions, it has been shown that data are sensitive to nuclear
correlations. In turn, these strictly depend on the NN potential. Therefore,
two-nucleon emission is a promising field deserving further investigation both
experimentally and theoretically in order to solve a longstanding problem in
nuclear physics. 

In particular, FSI's must be carefully treated. A consistent evaluation of FSI's
would require a genuine three-body approach, for the two nucleons and the
residual nucleus, by summing up an infinite series of contributions  in the
NN-scattering amplitude and in the interaction of the two nucleons with the
residual nucleus (Fig.~\ref{fig:fig1}). So far, only the major contribution of
FSI's, due to the interaction of each of the two outgoing 
nucleons with the residual nucleus, was taken into account in the different
models. The guess was that the mutual interaction between the two outgoing
nucleons (NN-FSI) could be neglected since they are mainly ejected back to back.

Results have been presented here with a first estimate of the role of NN-FSI in
the case of the $^{16}$O(e,e$'$pp) and $^{16}$O(e,e$'$pn) reactions.
In general, the optical potential gives an overall and substantial reduction of
the calculated cross sections. This important effect represents the main
contribution of FSI's and can never be neglected. In  most of the situations
considered here, NN-FSI gives an enhancement of the cross section. The effect 
is in general non negligible, it depends strongly on the
kinematics~\cite{schwamb1}, on the type of reaction~\cite{schwamb2}, and on the
final state of the residual nucleus~\cite{schwamb1}. NN-FSI affects in a
different way the 
various terms of the nuclear current, usually more the two-body than the
one-body terms, and is sensitive to the various theoretical ingredients of the
calculation. This makes it difficult to make predictions about the role of
NN-FSI in a particular situation. In general each specific situation should be
individually investigated. 

Extension of the same NN-FSI approach to real-photon-induced reactions will be
presented elsewhere~\cite{schwamb2}.
The solution of the full three-body problem of the final state is in progress.

\bc
{\bf Acknowledgements}
\ec

This work has partly been performed under the contract HPRN-CT-2000-00130 of the
European Commission. Moreover, it has been supported by the Istituto Nazionale
di Fisica Nucleare and by the Deutsche Forschungsgemeinschaft (SFB 443).




\begin{thebibliography}{99}


\bibitem{Oxford}
S.~Boffi, C.~Giusti, F.D.~Pacati and M.~Radici, {\em Electromagnetic Response of
Atomic Nuclei}, Clarendon Press, Oxford, 1996.

\bibitem{Report}
S.~Boffi, C.~Giusti and F.D.~Pacati, Phys. Rep. 226 (1993) 1.

\bibitem{Polls}
A.~Polls {\sl et al.}, Phys. Rev. C 55 (1997) 810.

\bibitem{Udias99}
J.M.~Ud\'{\i}as {\sl et al.},  Phys. Rev. Lett. 83 (1999) 5451.

\bibitem{Kelly99}
J.J.~Kelly, Phys. Rev. C 60 (1999) 044609.

\bibitem{Lapikas}
L.~Lapik\'as {\sl et al.}, Phys. Rev. C 61 (2000) 064325.

\bibitem{deForest}
T.~de Forest, Nucl. Phys. A 392 (1983) 232.

\bibitem{Kelly}
J.J.~Kelly, Phys. Rev. C 56 (1997) 2672.

\bibitem{Udias}
J.M.~Ud\'{\i}as {\sl et al.}, Phys. Rev. C 64 (2001) 024614.

\bibitem{meucci02}
A.~Meucci, Phys. Rev. C 65 (2002) 044601.

\bibitem{Meucci}
A.~Meucci, C.~Giusti and F.D.~Pacati, Phys. Rev. C 64 (2001) 014604.

\bibitem{rad-roth}
M.~Radici, W.H.~Dickhoff, E.~Roth Stoddard, Phys. Rev. C 66 (2002) 014613.

\bibitem{rad-dick}
M.~Radici, A.~Meucci, W.H.~Dickhoff, Eur. Phys. J. A 17 (2003) 65.

\bibitem{gurts}
W.J.W.~Geurts {\sl et al.}, Phys. Rev. C 53 (1996) 2207.

\bibitem{Barbieri}
C.~Barbieri and W.H.~Dickhoff, Phys. Rev. C 63 (2001) 034313; Phys. Rev. C 65
(2002) 064313.

\bibitem{Ulmer}
P.E.~Ulmer {\sl et al.}, Phys. Rev. Lett. 59 (1987) 2259; R.W.~Lourie {\sl et
al.}, Phys. Rev. C 57 (1993) R444; M.~Holtrop {\sl et al.}, Phys. Rev. C 58
(1998) 3205.

\bibitem{Dutta}
D. Dutta {\sl et al.}, Phys. Rev. C 61 (2000) 061602(R). 

\bibitem{Liyanage}
N. Liyanage {\sl et al.} (E-89-003),  Phys. Rev. Lett. 86 (2001) 5671.

\bibitem{schwamb1}
M.~Schwamb, S.~Boffi, C.~Giusti and F.D.~Pacati, Eur. Phys. J. A 17 (2003) 7.

\bibitem{schwamb2}
M.~Schwamb, S.~Boffi, C.~Giusti and F.D.~Pacati, nucl-th/0307003.

\bibitem{Gottfried}
K.~Gottfried, Nucl. Phys. 5 (1958) 557.

\bibitem{McG95}
J.C.~McGeorge {\sl et al.}, Phys. Rev. C 51 (1995) 1967.

\bibitem{Lam96}
Th.~Lamparter {\sl et al.}, Z. Phys. A 355 (1996) 1.

\bibitem{McG98}
J.D.~McGregor {\sl et al.}, Phys. Rev. Lett. 80 (1998) 245.

\bibitem{Wat00}
D.P.~Watts {\sl et al.}, Phys. Rev. C 62 (2000) 014616.

\bibitem{Onderwater97}
C.J.G.~Onderwater {\sl et al.}, Phys. Rev. Lett. 78 (1997) 4893.

\bibitem{Onderwater98}
C.J.G.~Onderwater {\sl et al.}, Phys. Rev. Lett. 81 (1998) 2213.

\bibitem{Starink}
R.~Starink  {\sl et al.}, Phys. Lett. B 474 (2000) 33.

\bibitem{Rosner}
G.~Rosner, Prog. Part. Nucl. Phys. 44 (2000) 99.

\bibitem{Boffi}
S.~Boffi, in {\sl Two Nucleon Emission Reactions}, ed. by O.~Benhar and
A.~Fabrocini (ETS Editrice, Pisa, 1990), p. 87.

\bibitem{GP}
C.~Giusti and F.D.~Pacati, Nucl. Phys. A 615 (1997) 373.

\bibitem{Giusti98}
C.~Giusti {\sl et al.}, Phys. Rev. C 57 (1998) 1691.

\bibitem{Geurts}
W.J.W.~Geurts {\sl et al.}, Phys. Rev. C 54 (1996) 1144.

\bibitem{Ryckebusch}
J.~Ryckebusch {\sl et al.}, Phys. Lett. B 441 (1998) 1.

\bibitem{A1-5-98}
A1/5-98: P.~Grabmayr and G.~Rosner, spokesmen.

\bibitem{GP99}
C.~Giusti {\sl et al.}, Phys. Rev. C 60 (1999) 054608.

\bibitem{GP00}
C.~Giusti and F.D.~Pacati, Phys. Rev. C 61 (2000) 054617.


\bibitem{GP01}
C.~Giusti and F.D.~Pacati, Eur. Phys. J. A 12 (2001) 69.

\bibitem{Nad81}
A.~Nadasen {\sl et al.}, Phys. Rev. C 23 (1981) 1023.


\end{thebibliography}
\end{document}